\documentclass[reprint,prb,superscriptaddress,amsmath,amssymb,aps,floatfix,longbibliography]{revtex4-2}
\usepackage{graphicx}
\usepackage{bm}
\usepackage{dcolumn}
\usepackage{hyperref}
\usepackage{float}
\usepackage{siunitx}
\usepackage[caption = false]{subfig}
\usepackage{xcolor}
\usepackage{textcomp}
\usepackage{gensymb}

\let\revappendix\appendix

\begin{document}

\title{Long distance electron-electron scattering detected with point contacts}

\author{Lev V. Ginzburg}
\email{glev@phys.ethz.ch}
\author{Yuze Wu}
\author{Marc P. R\"o\"osli}
\author{Pedro Rosso Gomez}
\author{Rebekka Garreis}
\author{Chuyao Tong}
\affiliation{Solid State Physics Laboratory, ETH Z\"urich, CH-8093 Z\"urich, Switzerland}
\author{Veronika Star\'{a}}
\affiliation{CEITEC - Central European Institute of Technology, Brno University of Technology, Purkyňova 123, 612 00 Brno, Czech Republic}
\author{Carolin Gold}
\affiliation{Solid State Physics Laboratory, ETH Z\"urich, CH-8093 Z\"urich, Switzerland}
\affiliation{Department of Physics, Columbia University, New
York, NY, USA}
\author{Khachatur Nazaryan}
\affiliation{Department of Physics, Massachusetts Institute of Technology, Cambridge, MA 02139, USA}
\author{Serhii Kryhin}
\affiliation{Department of Physics, Harvard University, Cambridge, MA 02138, USA}
\author{Hiske Overweg}
\author{Christian Reichl}
\author{Matthias Berl}
\affiliation{Solid State Physics Laboratory, ETH Z\"urich, CH-8093 Z\"urich, Switzerland}

\author{Takashi Taniguchi}
\affiliation{International Center for Materials Nanoarchitectonics, National Institute for Materials Science, 1-1 Namiki, Tsukuba 305-0044, Japan}
\author{Kenji Watanabe}
\affiliation{Research Center for Functional Materials, National Institute for Materials Science, 1-1 Namiki, Tsukuba 305-0044, Japan}
\author{Werner Wegscheider}
\author{Thomas Ihn}
\author{Klaus Ensslin}
\affiliation{Solid State Physics Laboratory, ETH Z\"urich, CH-8093 Z\"urich, Switzerland}

\date{\today}

\begin{abstract}
    We measure electron transport through point contacts in an electron gas in AlGaAs/GaAs heterostructures and graphene for a range of temperatures, magnetic fields and electron densities. We find a magnetoconductance peak around $B=0$. With increasing temperature, the width of the peak 
    increases monotonically, while its amplitude first increases and then decreases. For GaAs point contacts the peak is particularly sharp at relatively low temperatures $T\approx\SI{1.5}{K}$: the curve rounds on a scale of few tens of $\SI{}{\mu T}$ hinting at length scales  of several millimeters for the corresponding scattering processes. We propose a model based on the transition between different transport regimes with increasing temperature: from ballistic transport to few electron-electron scatterings to hydrodynamic superballistic flow to hydrodynamic Poiseuille-like flow. The model is in qualitative and, in many cases, quantitative agreement with the experimental observations.
\end{abstract}

\maketitle

\section{Introduction}

Electron transport can often be described by a semi-classical picture of charged particles moving through a material and interacting with impurities, phonons and sample boundaries. The two widely used models of electron flow - ballistic and diffusive (ohmic) - correspond to two opposite limits within this picture.  Ballistic transport usually describes the situation with few impurities and phonons, so that electrons mostly scatter with the sample boundaries, while the diffusive flow represents the case where momentum relaxation occurs mostly in the bulk of the system.

This picture changes considerably, if electron--electron scattering becomes significant. In clean systems, where the electron--electron mean free path $l_{ee}$ is much shorter than both the characteristic sample size and the transport mean-free path $l_\tau$, electron transport is similar to viscous flow of a classical fluid. This is known as the viscous (or hydrodynamic) electron transport regime \cite{Gurzhi63}.

Viscous electron flow was observed in different materials, including GaAs\cite{Molenkamp95, Braem2018, Ginzburg2021, Keser2021}, graphene\cite{Bandurin2016, KrishnaKumar2017, Berdyugin2019}, PdCoO$_2$ \cite{Moll2016}, WP$_2$ \cite{Gooth2018} and WTe$_2$ \cite{Vool2021}. Experimental evidence for hydrodynamic behavior comes from superballistic flow through point contacts\cite{KrishnaKumar2017, Ginzburg2021}, negative nonlocal resistance \cite{Bandurin2016, Braem2018}, the Gurzhi effect \cite{Molenkamp95}, Stokes flow \cite{Gusev2020}), scanning probe experiments investigating Poiseuille flow \cite{Sulpizio2019, Ku2020, Vool2021, Krebs2021}. Most experiments were performed at zero magnetic field. Magnetic fields high enough, such that the cyclotron radius becomes the shortest relevant length scale in the system, will eventually eliminate hydrodynamic effects \cite{Alekseev16}. However, the intermediate regime of small magnetic fields offers an interesting playground where several length scales 
compete. A magnetic field introduces a Lorentz force acting on the electron system. Furthermore, it modifies the viscosity and adds a second viscosity coefficient, usually called the Hall viscosity \cite{Avron1998, Alekseev16, Pellegrino2017}. The interplay of viscous flow and magnetic field was experimentally investigated in a vicinity geometry \cite{Berdyugin2019} as well as in channels wider than the electron--electron mean free path \cite{Gusev2018_1, Gusev2018_2}.

In this paper we focus on one of the simplest possible structures in two-dimensional electron gases (2DEG) - point contacts (PCs). Electron transport through the PCs at low temperatures is well-understood and in general can be explained well within ballistic approximations. It was shown before that at higher temperatures hydrodynamic effects begin to play a significant role: at zero magnetic field the conductance exceeds the fundamental ballistic (Sharvin) limit due to the collective movement of electrons reducing momentum loss. This effect was predicted theoretically \cite{Guo2016} and observed experimentally in graphene \cite{KrishnaKumar2017} and GaAs \cite{Ginzburg2021}.

Non-zero magnetic fields add further complexity to the system: a peak in the magnetoconductance is observed around zero magnetic field at elevated temperatures ($\approx\SI{10}{K}$ in GaAs PCs). This peak was observed previously in GaAs PCs \cite{Renard2008, Melnikov2012} and was associated with  electron--electron interactions. Here we expand on these early observations for PCs in GaAs two-dimensional electron gases (2DEGs) and investigate the non-monotonic behavior of the magnetoconductance peak as a function of temperature, carrier density and split-gate voltages. Peculiar properties of the conductance peak that were previously unnoticed include extreme sharpness of the peak (rounding on a scale of few tens of $\SI{}{\micro T}$) at relatively low temperatures ($T<\SI{2}{K}$) and slow disappearance of the peak at high temperatures ($T>\SI{10}{K}$). Furthermore, we demonstrate that some of these effects can be observed in graphene PCs at $\approx\SI{100}{K}$ which is the temperature range where viscous flow occurs in graphene \cite{KrishnaKumar2017}.

We argue that the observations can be explained by a continuous transition between different transport regimes with increasing temperature. At very low temperatures ($T<\SI{1}{K}$ for GaAs PCs) transport is mostly ballistic. At $\SI{1}{K}<T<\SI{2}{K}$ electron-electron interactions become more important and the system is not ballistic, but also not yet fully hydrodynamic. This transitional regime results in a small but very sharp peak. At higher temperatures $l_\tau \gg l_{ee}$ and electron transport becomes hydrodynamic. For $T<\SI{10}{K}$ the PC width $d$ remains small compared to the other relevant length scales. Electron transport is superballistic, the peak in the magnetoconductance is present. At even higher temperatures $d$ becomes comparable or smaller than the electron-electron mean-free path, and electron transport starts to resemble Poiseuille flow. As a consequence the peak in conductance becomes less pronounced and eventually almost disappears. 
For graphene PCs the relevant temperatures are generally 
higher, and we observe only ballistic and superballistic regimes. The model that we present agrees with our observations qualitatively and, in most cases, quantitatively.


\section{Methods}

In this paper we use both GaAs and graphene devices. 

The first device is based on a AlGaAs/GaAs heterostructure with a 2DEG $\SI{200}{n m}$ below the surface. The global patterned back-gate allows us to change the electron density  between $\SI{1.5e11}{c m^{-2}}$ and $\SI{2.7e11}{c m^{-2}}$ \cite{Berl2016}. The low-temperature (below $\SI{1}{K}$) mobility is up to $\SI{7e6}{cm^{2}/Vs}$ corresponding to a transport mean-free path of more than $\SI{60}{\micro m}$. The device has a shape of a large multi-terminal Hall bar ($\SI{1800}{\micro m} \times \SI{400}{\micro m}$) with several top-gate defined PCs (lithographic width $d=\SI{250}{n m}$) in the central part of the Hall bar.

The second device comprises of monolayer graphene encapsulated between hexagonal boron nitride crystals (hBN) with a graphite back-gate. The stack is made with the standard dry-transfer technique and is placed on top of a silicon chip with $\textrm{SiO}_2$ surface. The Hall bar shape of the device and the PCs are etched through the top hBN with reactive ion etching. The widths of the PCs are $d_\mathrm{narrow} = \SI{150}{n m}$ and $d_\mathrm{wide} = \SI{350}{n m}$.

All linear conductance measurements were performed in a $^4$He and $^4$He/$^3$He systems at temperatures between $\SI{0.25}{K}$ and $\SI{25}{K}$ for the GaAs device and between $\SI{4.2}{K}$ and $\SI{120}{K}$ for the graphene device. Standard lock-in techniques at 31 Hz were used. The carrier densities $n$ were measured using the classical Hall effect (GaAs) and Shubnikov–de Haas oscillations (graphene). The magnetic field $B$ is always perpendicular to the surface of the sample. All measurements are 4-terminal. For the GaAs device, only one pair of top-gates was used at a time with all other top-gates grounded.

\section{Measurements}

First, measurements in the GaAs device were performed. Below the data for one PC are presented; the data for other PCs were consistent with these observations and displayed similar features (see Appendix).

Figure~\ref{fig:1}(a) shows an example curve for PC conductance $G$ as a function of $B$ at $T=\SI{1.3}{K}$ and $n=\SI{2.7e11}{c m^{-2}}$. The overall behavior is well known: the conductance increases linearly with $|B|$ \cite{vanHouten1988_1} until the onset of Shubnikov–de Haas oscillations, which for this sample is visible above $|B|>\SI{200}{mT}$. A peak in conductance is present around $B=0$.

First, we focus on a relatively narrow temperature range between $\SI{1.3}{K}$ and $\SI{4.3}{K}$ (Figure~\ref{fig:1}(b)). The magnetoconducatance peak becomes more pronounced with increasing temperature, and less pronounced but sharper with decreasing temperature. Figure~\ref{fig:1}(c) shows $\Delta G(B,T) = G(B,T) - G(0,T)$ as a function of $B$ in a narrow range of $B$. Note that at low temperature $T=\SI{1.3}{K}$ the peak is rounded on a scale of few tens of $\SI{}{\micro T}$. Features of this size in magnetic field are in general highly unusual in electron transport, and to the best of our knowledge were not reported for AlGaAs/GaAs heterostructures before.

\begin{figure*}
    \includegraphics[width=\linewidth]{{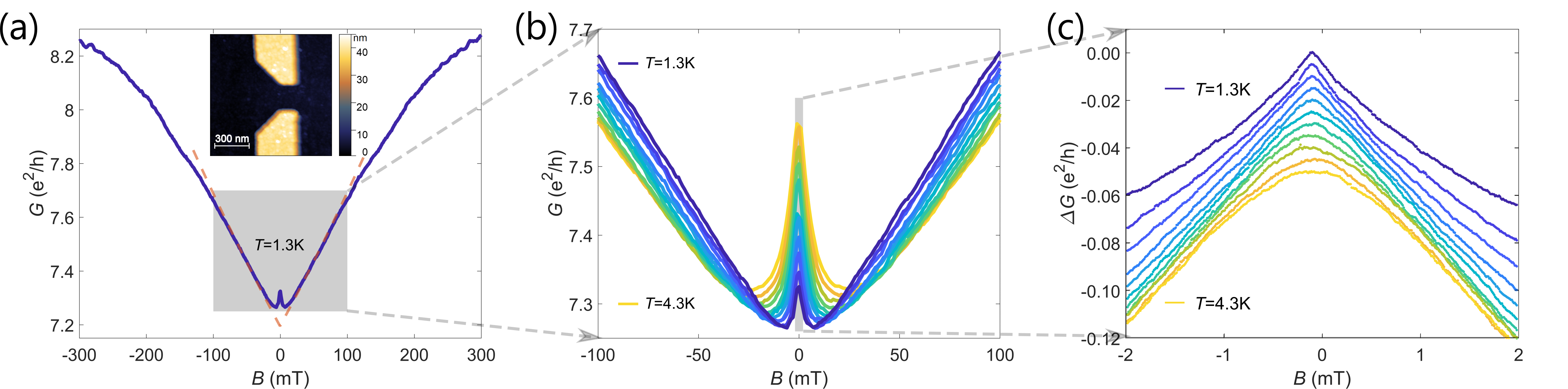}}
    \caption{(a) Conductance of a GaAs PC  as a function of magnetic field $G(B)$ at electron density $n=\SI{2.7e11}{c m^{-2}}$ and temperature $T=\SI{1.3}{K}$. The inset shows an AFM image of the PC. The range close to $B=0$ (gray area) is shown in more detail and for temperatures between $T=\SI{1.3}{K}$ and $T=\SI{4.3}{K}$ with a step of $\Delta T=\SI{0.3}{K}$ in (b). In (c) we further zoom into a very small range of $B$. Conductance with respect to the peak value for each temperature $\Delta G(B,T) = G(B,T) - G(0,T)$ is shown. The curves in (c) are vertically offset for clarity.}
    \label{fig:1}
\end{figure*}

\begin{figure*}
    \includegraphics[width=\linewidth]{{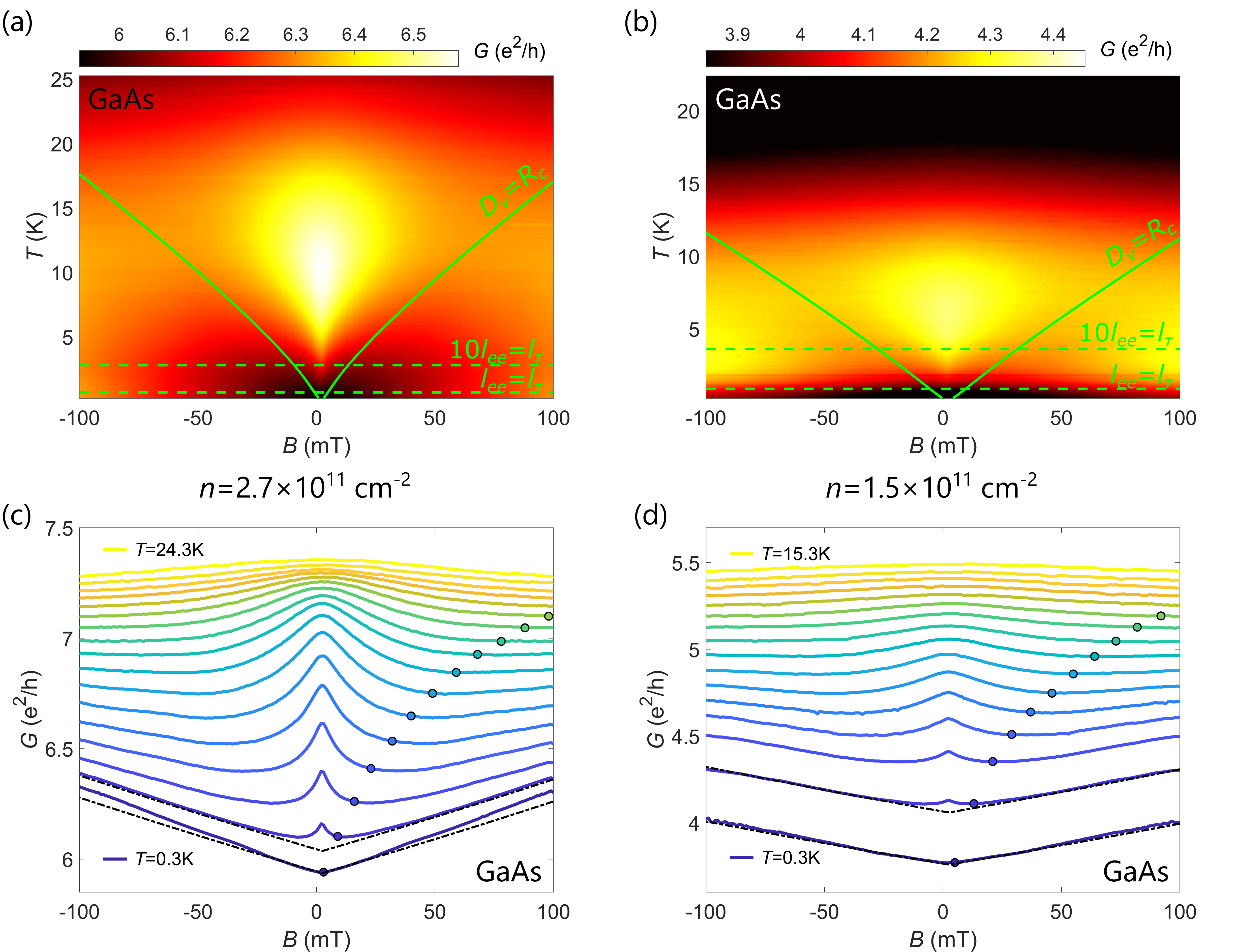}}
    \caption{Conductance of a GaAs PC $G$ as a function of magnetic field $B$ and temperature $T$. (a) and (c) correspond to the electron density $n=\SI{2.7e11}{c m^{-2}}$, (b) and (d) correspond to electron density $n=\SI{1.5e11}{c m^{-2}}$. The full data set is presented in 2D-maps (a) and (b). The solid green lines represent $D_\nu=R_\mathrm{C}$ (above the lines $D_\nu<R_\mathrm{C}$, below $D_\nu>R_\mathrm{C}$). The upper green dashed line represents $l_\tau=10l_{ee}$, the lower one $l_\tau=l_{ee}$. (c) and (d) show line cuts at constant temperatures of the (a) and (b) plots respectively. The step in temperature between two lines is $\SI{1.5}{K}$ for (c) and $\SI{1}{K}$ for (d). The lines are shifted artificially by $\SI{0.075}{e^2/h}$ for (c) and $\SI{0.1}{e^2/h}$ for (d). Large dots on the lines correspond to the magnetic field value where $D_\nu=R_\mathrm{C}$ (only shown for positive $B$). The dashed-dotted black lines show the ballistic fit (suppressed backscattering) for the two lowest temperatures. Calculations of $D_\nu$ and $R_\mathrm{C}$ use no fitting parameters; the ballistic fit calculations use $G(B=B_0)$ as the only fitting parameter (here $B_0 = \SI{2.5}{m T}$ is the effective zero of the magnetic field).}
    \label{fig:2}
\end{figure*}

Second, we explore the properties of the peak in a wider temperature range. Figures~\ref{fig:2}(a, b) show the conductance $G$ of the PC as a function of $B$ and $T$ for two different electron densities $n=\SI{2.7e11}{c m^{-2}}$ (a) and $\SI{1.5e11}{c m^{-2}}$ (b). For the higher electron density a small step in conductance occurred at $T=\SI{13.6}{K}$, which we attribute to a random impurity being charged/discharged close to the PC; in order to compensate for it, the conductance above this temperature is multiplied by $0.997$.

The background increase of conductance with $|B|$, mentioned above, is clearly present at low temperatures and becomes less pronounced with increasing temperature. The peak in conductance around $B=0$ is present at all $T>\SI{0.5}{K}$ and is most visible around $T\approx\SI{12}{K}$.

The amplitude of this peak first increases and then decreases with temperature, with the maximum being around $\approx\SI{10}{K}$ for all measured electron densities. The maximum is placed at slightly higher temperatures for higher electron densities. The width of the peak along the magnetic field axis increases monotonically with temperature for all densities. The details of the $G(B)$ can be seen as line cuts at constant temperatures in Figures~\ref{fig:2}(c, d).

A similar behavior was observed for all available electron densities in several PCs in this device as well as in a PC in a different GaAs device (no back-gate, fixed electron density $\SI{1.8e11}{c m^{-2}}$, mobility up to $\SI{4.1e6}{cm^{2}/Vs}$, PC width $d=\SI{500}{n m}$, see Appendix). For as long as the PC was defined, the top-gate voltage $V_\mathrm{TG}$ and therefore the effective width of the PC affected the amplitude of the peak, but not its width along the magnetic field axis. Some indications of the conductance peak are present even in a very wide PC (lithographic width $d=\SI{4}{\micro m}$, see Appendix B for details).

\begin{figure*}
    \includegraphics[width=\linewidth]{{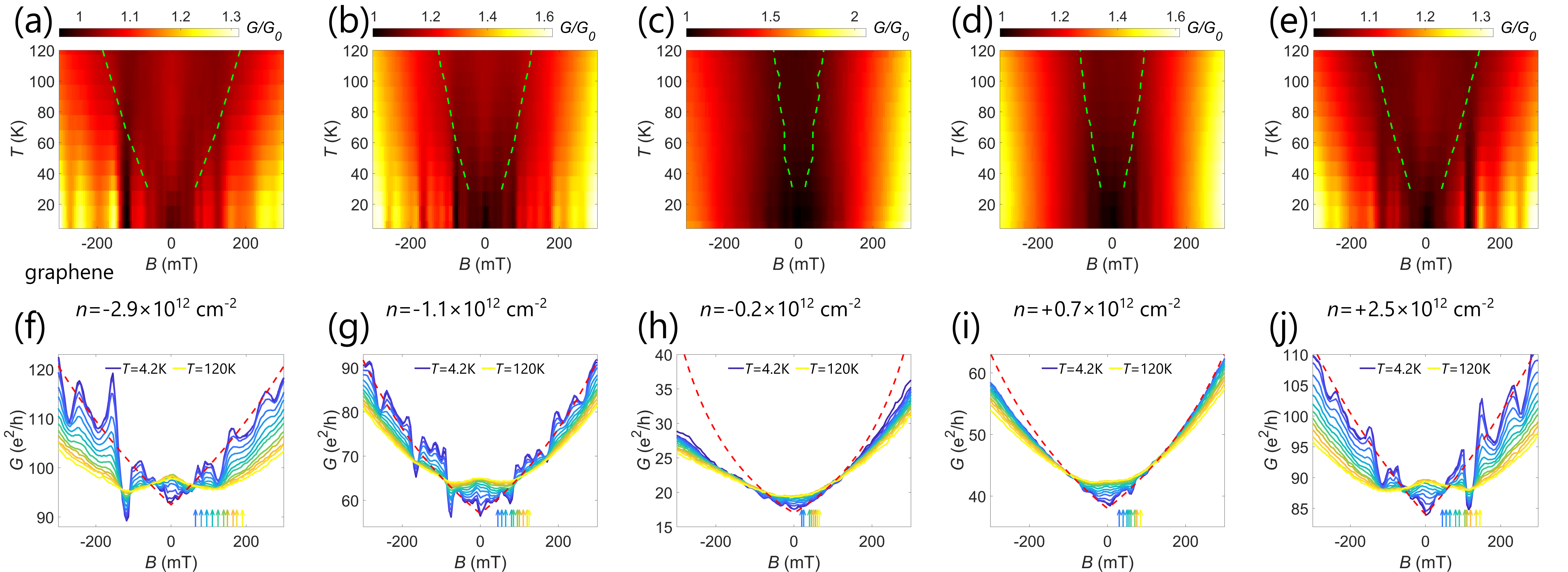}}
    \caption{Conductance of graphene PC $G$ as a function of magnetic field $B$ and temperature $T$ at different carrier densities. Negative carrier densities correspond to holes, positive ones correspond to electrons. (a) - (e) 2D maps with green dashed lines representing the magnetic field values for which $D_\nu=R_\mathrm{C}$. Conductance is normalized by $G_0 = G(B=0, T=\SI{4.2}{K})$ for each plot. (f) - (j) Line cuts at constant temperatures (starting at $\SI{4.2}{K}$, then from $\SI{10}{K}$ to $\SI{120}{K}$ with a step of $\SI{10}{K}$). Magnetic field $B$ such that $D_\nu=R_\mathrm{C}$ is shown for every temperature by an arrow of corresponding color. Red dashed lines show the ballistic fit (suppressed backscattering). Calculations of $D_\nu$ and $R_\mathrm{C}$ use no fitting parameters; ballistic calculations use $G(B=0)$ as the only fitting parameter.}
    \label{fig:3}
\end{figure*}

Similar, but quantitatively different effects were observed in graphene PCs (Figure~\ref{fig:3}) at temperatures $\approx\SI{100}{K}$. Figures~\ref{fig:3}(a - e) show the  normalized conductance $G(B,T)$ for different carrier densities $n$ (positive sign of $n$ corresponds to electrons) with corresponding linecuts presented in Figues~\ref{fig:3}(f - j). At low temperatures we observe an increase in conductance with increasing $|B|$ due to suppressed backscattering \cite{vanHouten1988_1}. Unlike for GaAs PCs, there are oscillations superimposed on the V-shaped background. These oscillations are not symmetric in $B$ and decay with temperature. Fast Fourier transformation of the curves provides a leading period in magnetic field proportional to $\sqrt{|n|}$ and the corresponding cyclotron radius $R_\mathrm{C}\approx\SI{2.5}{\micro m}$; these oscillations likely result from magnetic focusing between the PCs and the narrow voltage probes.

Similar to GaAs PCs, we observe an increase of conductance around $B=0$ with increasing temperature. This behavior is more pronounced and visible as a peak in $G(B)$ for high hole or electron densities (e.g. $\SI{-2.9e12}{c m^{-2}}$ for holes and $\SI{2.5e12}{c m^{-2}}$ for electrons, Figures~\ref{fig:3}(f, j)). At lower carrier densities the peak is not present, but there is still a noticeable increase in $G(B)$ around zero magnetic field at higher temperatures (Figures~\ref{fig:3}(h, i)).

In contrast to GaAs PCs, the conductance peak for graphene PCs does not become very sharp at any measured temperatures or carrier densities, i.e. the rounding of the peak always happens on a scale of tens of $\SI{}{m T}$.

\section{Discussion}

\subsection{Hydrodynamic Model}

In this section we discuss possible explanations for the conductance peak behavior as shown in Figure~\ref{fig:2} and propose a quantitative model. 

To begin with, we exclude possible explanations based on a single-particle picture. The conductance peak is not a classical size effect, since the width of the peak depends significantly on temperature, while the cyclotron radius does not. Neither is it caused by weak antilocalization, as here the effect becomes stronger with increasing temperature (at least up to a certain temperature) and is not present in bulk measurements. The observed effect is also not caused by the filling of a second subband of the 2DEG: while this could produce a peak of conductance around $B=0$ \cite{vanHouten1988_2}, and the amplitude of this peak can theoretically increase with temperature if the occupation of the second subband increases, this effect should have a strong dependence on electron density, which is not the case here. No other single-particle effects seem to be able to produce the observed behavior.

Both the temperature range in which the effect is observed and the non-monotonic temperature dependence suggest an explanation related to electron-electron interactions. Below we show that the viscous electron transport model can indeed explain our observations.

First, let us consider $B=0$. In this case the PC conductance can be written as the sum of two terms (the so-called superballistic electron flow model \cite{Guo2016}): 
\begin{equation}
    G=G_\mathrm{ball}+G_\mathrm{vis}
    \label{eq:sum}
\end{equation}
The first part $G_\mathrm{ball}$ is the same as expected for purely ballistic electron transport (Sharvin conductance \cite{Sharvin1965}) $G_\mathrm{ball} = Ng_\mathrm{v}g_\mathrm{s}\frac{e^2}{h}$, where $N$ is the number of modes in the PC, $g_\mathrm{s}$ is the spin degeneracy and $g_\mathrm{v}$ is the valley degeneracy. The second part $G_\mathrm{vis}$ is the viscous contribution, originating from the Stokes equation (low Reynolds number, no magnetic field, steady state approximations): 
\begin{equation}
    \vec{J} - D_\nu^2\nabla^2\vec{J} = -\sigma_0\nabla\phi,
    \label{eq:Stokes}
\end{equation}
where $\sigma_0$ is the ohmic ``Drude-like'' conductance due to impurities and phonons, $\phi$ is the electrostatic potential, $\vec{J}$ is the current density and  
\begin{equation}
    D_\nu=\frac{\sqrt{l_{ee}l_\tau}}{2}
    \label{eq:dnu}
\end{equation}
is the length scale responsible for viscous flow (also described as vorticity diffusion length) \cite{Torre2015}.

At high enough magnetic field, such that the cyclotron radius $R_\mathrm{C} = \frac{\hbar k_\mathrm{F}}{eB}$ is much shorter than the length scale for the viscous transport, hydrodynamic effects are not present and we recover the ballistic result. The conductance increases with increasing $|B|$ due to the suppression of backscattering as $G_\mathrm{ball} = \frac{1}{R_0 - |B|/en} \approx G_0 + G_0^2\frac{|B|}{en}$ \cite{vanHouten1988_1} with possible oscillations on top due to the Shubnikov-de Haas effect (here $G_0 = G_\mathrm{ball}(B=0)$). Higher temperatures remove the oscillations and reduce the dependence of conductance on the magnetic field due to electron-phonon scattering. If we use this formula for $G_\mathrm{ball}$, as well as extend the definition of $G_\mathrm{vis}$ to non-zero magnetic field so that it is zero in high $B$, Eq. \ref{eq:sum} becomes applicable both in zero and in high $B$.

The applicability of equation~\ref{eq:sum} in intermediate fields is not known; however, if we assume a smooth transition between zero $B$ and high $B$ regions, we arrive at a result that fits our observations. Indeed, we experimentally observed a V-shaped background in $G(B)$ that becomes less pronounced at higher temperatures, which corresponds to the results of ballistic calculations (shown as dashed-dotted black lines in Figures~\ref{fig:2} (c,d)). On top of this, the viscous contribution $G_\mathrm{vis}$ creates a conductance peak centered around $B=0$ (see Figure~\ref{fig:3} (a)). This peak grows with increasing temperature as $l_{ee}$ and $l_\tau$ become shorter (see Appendix A for the temperature dependence of the length scales). Increasing the temperature further (such that optical phonons become relevant) would eventually destroy both ballistic and hydrodynamic effects and recover purely diffusive behavior.

Next, we discuss the width of the conductance peak along the magnetic field axis. A magnetic field introduces an additional Lorentz-like term to the Equation~\ref{eq:Stokes} as well as modifies the viscous term; both additions have a corresponding length scale $R_\mathrm{C}$. To the best of our knowledge, the analytical solution to the resulting equation is not known, so we attempted to extract some estimates from the comparison of length scales. There are three different length scales in this problem: $R_\mathrm{C}$ from the magnetic field, $D_\nu$ from the original Stokes equation, and the width of the PC $d$ from the boundary conditions. $d$ does not depend on magnetic field or temperature and is not likely to give us the observed non-trivial behavior of the conductance peak. Therefore dimensionality dictates $R_\mathrm{C}\sim D_\nu$. Here we hypothesise that $R_\mathrm{C}=D_\nu$ is the transition point for electron behavior: in this case, for $R_\mathrm{C}>D_\nu$ some increase of conductance above the V-shaped background due to viscous effects would still present; for $R_\mathrm{C}<D_\nu$ the superballistic behavior would mostly be gone.

In order to compare this hypothesis with the experiment, we numerically calculate $l_{ee}$ \cite{Jungwirth1996} (see Appendix A for the details) and extract $l_\tau$ from the direct measurements of the bulk resistance (the large size of the Hall bar allows the use of the Drude model). Together these two mean-free paths give us the viscous length $D_\nu(T)$ (Equation~\ref{eq:dnu}). From the assumption $R_\mathrm{C}=D_\nu$ we solve $R_\mathrm{C}(B)=D_\nu(T)$ and get $B_\mathrm{C}(T)$ for the transition in electron behavior. The solid green lines in Figures~\ref{fig:2} (a, b) and large dots in Figures~\ref{fig:2}(c, d) correspond to $B(T) = \pm B_\mathrm{C}(T) + B_0$, where experimental offset $B_0 = \SI{2.5}{m T}$. The hydrodynamic model is only applicable when $l_\tau \gg l_{ee}$; in order to show the region of the model validity we add two green dashed lines to Figures~\ref{fig:2}(a, b): the lower ones correspond to temperatures where $l_\tau = l_{ee}$ and the upper ones correspond to $l_\tau = 10l_{ee}$. Indeed, we see that the observed peak of conductance lies between two solid green lines and above the dashed ones, i.e. in the region where $D_\nu(T)<R_\mathrm{C}(B)$ and the hydrodynamic model is applicable. The positions of $B(T) = \pm B_\mathrm{C}(T) + B_0$ closely correspond to the minima of $G(B,T)$. Consequently, the measurements support the hypothesis that $R_\mathrm{C}=D_\nu$ is the relevant condition.

The expected change of $B_\mathrm{C}(T)$ with experimental parameters is also consistent with observations. A lower electron density gives higher values of $B_\mathrm{C}$ at a given temperature, which correspond to a wider peak in Figure~\ref{fig:2}(b) compared to Figure~\ref{fig:2}(a).

Following the same approach as for GaAs, we compare the cyclotron radius $R_\mathrm{C}$ with the viscous length $D_\nu$ for graphene. We estimate $l_\tau$ from the bulk resistance measurements. Unfortunately, unlike in the GaAs device, the width of the Hall bar and the distance between the contacts in the graphene device is not large compared to $l_\tau$, therefore the Drude model based calculations of $l_\tau$ are not precise. The real transport mean-free path is likely longer than the calculated one. For the estimate of $l_{ee}$ we use the experimental and numerical results from \cite{KrishnaKumar2017}, where a similar device was used. We scale the $l_{ee}$ to the electron densities in our measurements according to $l_{ee} \sim \frac{\sqrt{n}}{ln(n)}$. The resulting $B_\mathrm{C}(T)$ is shown in Figure~\ref{fig:3}. The observed peak of conductance is somewhat narrower along the magnetic field axis than the calculated $B_\mathrm{C}$; this can be explained by the underestimated transport mean-free path $l_\tau$. Aside from this inconsistency, the experimental results for the graphene device correspond to the suggested model.

\subsection{Sharpness of the Conductance Peak}

The extreme sharpness of the conductance peak does not follow from the hydrodynamic model described above. In addition, not only the size, but also the shape of the peak changes significantly with temperature: at low temperatures, the peak is wide at the bottom and sharp at the top, while at high temperatures it has a more rounded shape. This change in shape hints at the presence of a second relevant length scale, in addition to $D_\nu$, with a temperature dependence different from $D_\nu$. This new length scale should be particularly large at low temperatures (up to a few $\SI{}{mm}$) in order to explain the observations of the sharp conductance peak at $T=\SI{1.3}{K}$ (Figure~\ref{fig:1}(c)).

The hydrodynamic model is based on the assumption of many electron-electron interactions before momentum is dissipated, i.e. $l_\tau \gg l_{ee}$. At $\SI{1}{K}<T<\SI{2}{K}$ however, this assumption is not valid anymore for GaAs: $l_\tau > l_{ee}$ but not $l_\tau \gg l_{ee}$, and the number of e-e scattering events before momentum is dissipated is not large. This transitional regime between ballistic and hydrodynamic transport potentially offers the additional length scale described above.

\begin{figure}
    \includegraphics[width=.8\linewidth]{{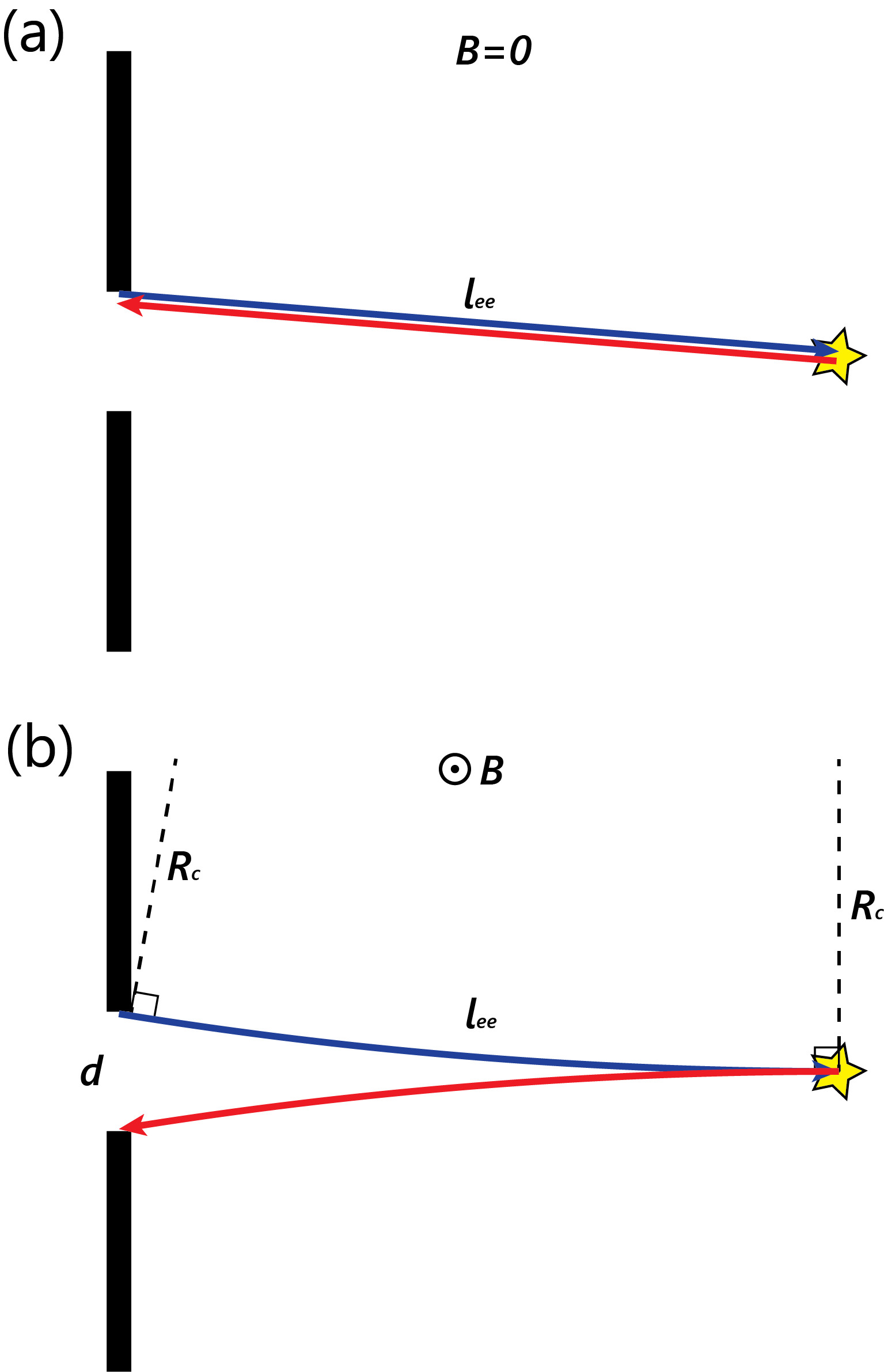}}
    \caption{Schematic of the electron back-scattering as a hole relevant at low temperatures ($\SI{1}{K}<T<\SI{2}{K}$) in zero (a) and non-zero (b) magnetic field. Black areas represents depleted parts of 2DEG forming the point contact. The electron moving through the PC is shown with blue arrow, and the back-scattered hole is shown with red arrow. The electron-electron scattering event happens at the position marked with yellow star.}
    \label{fig:4}
\end{figure}

Let us consider a single electron $e$ added above an equilibrium distribution, moving through a PC from left to right at $B=0$, represented by the blue arrow in Figure~\ref{fig:4}(a). After traveling for $\sim l_{ee}$ (several $\SI{}{\micro m}$ at $T\approx\SI{1.5}{K}$, see Appendix A for details), this electron scatters with another electron $e'$ from the equilibrium distribution (yellow star in the figure). Electron-electron scattering tends to be a head-on collision due to phase-space arguments \cite{Gurzhi1995, Ledwith2017, Kryhin2021}, i.e. $e'$ was moving towards the PC before the scattering event. 
In addition, $d \ll l_{ee}$ means that it is unlikely for any of the two electrons to move towards the PC after their scattering is small. Consequently, from the point of view of the total current through the PC, the most probable scattering event can be rephrased as an electron $e$ back-scattering as a hole $h$ towards the PC (red arrow in the Figure~\ref{fig:4}(a)), where $h$ represents the absence of $e'$ that had been moving towards the PC. In this case holes are defined not in a band structure meaning (as quasiparticles in the valence band) but as quasiparticles missing from the equilibrium distribution in the conductance band.

This back-scattered hole can go back through the PC and provide additional current, leading to a small increase in conductance. Multiple scattering events are also possible, i.e. an electron can back-scatter as a hole which can back-scatter again before reaching the PC as an electron, and so on. Here we only consider a single scattering event.

Let us add a small magnetic field $B$ perpendicular to the 2DEG. Now both the electron (before the scattering) and the back-scattered hole move in circular arcs, and their trajectories do not coincide anymore. Figure~\ref{fig:4}(b) shows an example case where the electron originates from one edge of the PC, and the resulting hole goes symmetrically through the PC at the other edge. In this case it can be geometrically derived that:

\begin{equation}
    R_\mathrm{C} \approx \frac{l_{ee}^2}{d}.
    \label{eq:sharp}
\end{equation}

Here we assumed $R_\mathrm{C} \gg l_{ee} \gg d$. Shorter $R_\mathrm{C}$ (higher magnetic field) would result in larger deviations between the trajectories of the electron and the hole, preventing the hole from going backwards through the PC and eliminating this additional contribution to conductance.

In a given temperature window Equation~\ref{eq:sharp} results in long $R_\mathrm{C}$ (up to several $\SI{}{m m}$), and consequently small $B$ ($\approx\SI{10}{\micro T}$). Below this value of $B$, the increase of conductance due to electron back-scattering is mostly unaffected by $B$; above this value, an increase in $B$ would cause some of the back-scattered holes to miss the PC and lead to a decrease in conductance. This can explain the sharpness of the conductance peak observed in the experiments.

The model is also consistent with a change in the peak shape with temperature. In leading order, $l_{ee}\sim T^{-2}$. Therefore the magnetic field describing sharpness of the peak scales as $B\sim T^4$. This temperature dependence is much stronger than the one following from the hydrodynamic model (Section IV.A): $B\sim D_\nu^{-1} \sim T^{3/2}$ or slower (here typical temperature dependence for electron-phonon scattering is used, $l_\tau \sim T^{-1}$, impurities will lead to saturation at low $T$. See also Appendix A). Consequently, the sharpness of the peak changes much faster with $T$ (at low temperatures) than the overall width of the peak (compare to Figure~\ref{fig:1}(b,c)).

The simplified argument above gives us an estimate of the length scale important for the problem. A complete model would have to account for multiple factors, including the angular distribution for the electrons arriving through the PC, different possible distances between the PC and the first scattering, multiple e-e scattering events, the finite possibility of momentum-relaxing scattering and the details of the electrostatic potential at the PC. Creating this detailed model goes beyond the scope of this paper.

Interestingly, we observe the sharp conductance peak only in GaAs devices and not in the graphene device. This can be explained by the smaller size of the graphene device, which is a technical limitation compared to GaAs: at low temperatures, where the described effect is important, an electron moving through the PC is more likely to scatter at a graphene flake boundary and not with another electron, suppressing the relevant contribution to the PC conductance at low magnetic field.

\subsection{Peak Amplitude}

\begin{figure*}
    \includegraphics[width=\linewidth]{{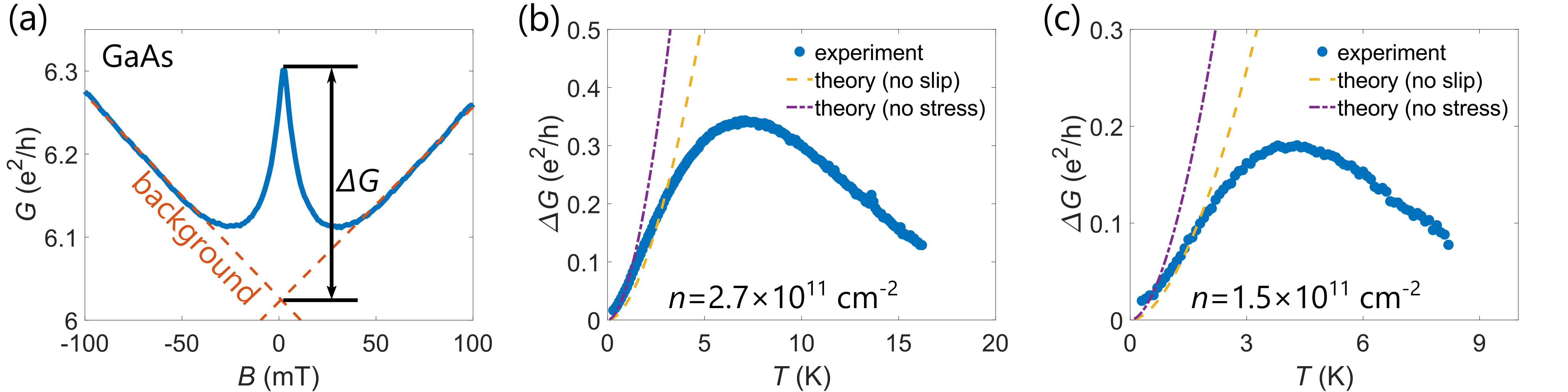}}
    \caption{Conductance peak height $\Delta G$ for GaAs PC. (a) Example of the $\Delta G$ calculation for one temperature ($\SI{7}{K}$) and electron density ($\SI{2.7e11}{c m^{-2}}$): the blue curve is the conductance $G$ as a function of $B$ at temperature $\SI{7}{K}$ and electron density $\SI{2.7e11}{c m^{-2}}$. The red dashed lines are linear fits at high magnetic field, $\Delta G$ is the excess conductance above the linear fits. $\Delta G$ calculated this way is shown as a function of temperature for two densities in (b) and (c)  (blue dots). Hydrodynamic calculations of superballistic conductance contributions at zero magnetic field are shown by yellow dashed lines for no slip and purple dashed-dotted lines for no stress boundary conditions.}
    \label{fig:5}
\end{figure*}

In the previous section we examined the transition regime between hydrodynamic and ballistic electron transport at low temperatures. Here we show that at higher temperatures there is another change of the system's behavior, not predicted by the hydrodynamic superballistic flow model. This can be seen by analyzing the temperature dependence of the amplitude of the conductance peak.

Following Equation~\ref{eq:sum}, which is applicable for superballistic flow model, the measured conductance is split into the background $G_\mathrm{ball}$ and viscous $G_\mathrm{vis}$ contributions. For each temperature the V-shaped background is approximated as a linear fit from the high magnetic field data (see red dashed lines in Figure~\ref{fig:5}(a) as an example). This fit is subtracted from $G(B)$ to acquire the zero $B$ viscous contribution $\Delta G$ as the height of the conductance peak. $\Delta G$ is shown as a function of temperature for two different electron densities in Figures~\ref{fig:5}(b, c). $\Delta G$ is not presented for the entire temperature range, since at higher temperatures the linear parts of $G(B)$ are not visible and the linear fit is not well defined. Next, the theoretically predicted viscous conductance contribution at $B=0$ is calculated (superballistic conductance) \cite{Guo2016}:

\begin{equation}
    G_\mathrm{vis}^\mathrm{th}=\frac{e^2d_\mathrm{eff}^2}{8\hbar}\sqrt{\frac{\pi n}{2}}\frac{1}{l_{ee}},
    \label{eq:Gvis}
\end{equation}

where $n$ is the bulk electron density and $d_\mathrm{eff}$ is the effective width of the PC. In general, $d_\mathrm{eff}$ is smaller than the lithographic width $d$ due to the side depletion below top-gates. Since the dependence of $d_\mathrm{eff}$ on temperature is weak, it can be calculated in the ballistic case from the lowest temperature zero $B$ value of $G$:

\begin{equation}
    d_\mathrm{eff}=\frac{h}{2e^2}\sqrt{\frac{\pi}{2n_\mathrm{PC}}}G\Big|_{T=T_\mathrm{min}, B=B_0}
    \label{eq:deff}
\end{equation}

The electron density in the PC $n_\mathrm{PC}$ present in the equation above is significantly lower than the bulk density $n$. We estimate it by comparing the Hall voltage and the diagonal voltages in the quantum Hall effect measurement. The result of the calculations is shown in Figures~\ref{fig:5}(b,c) as yellow dashed lines. However, equation \ref{eq:Gvis} is derived under no-slip boundary conditions, which is likely not the case for the gate-defined GaAs PCs. A paper by Li et al \cite{Li2021} predicts that for the no-stress boundary condition $G_\mathrm{vis}$ should be two times higher (purple dashed-dotted lines in Figures~\ref{fig:5}(b, c)).

\begin{figure*}
    \includegraphics[width=\linewidth]{{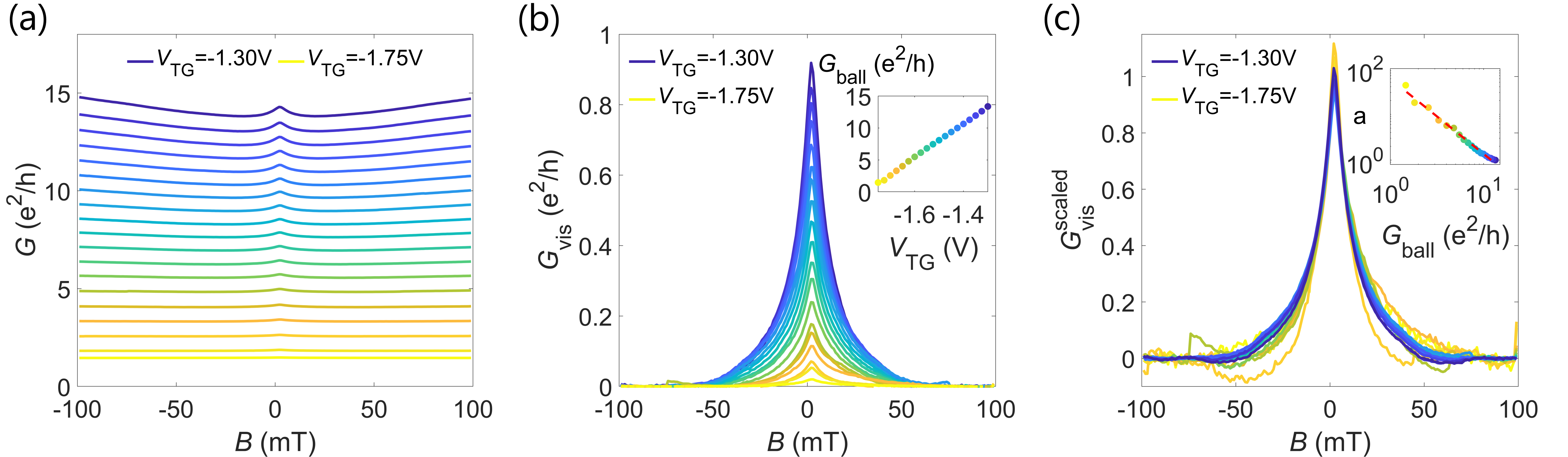}}
    \caption{(a) Conductance of GaAs PC $G$ as a function of magnetic field $B$ for different top-gate voltages $V_{TG}$. The temperature is fixed at $\SI{4.2}{K}$, the electron density is $\SI{2.7e11}{c m^{-2}}$. (b) Conductance peak as a function of $B$ for different $V_{TG}$, background V-shaped curve subtracted as in Figure~\ref{fig:5}. The inset shows the background conductance at effective zero magnetic field (for example, see the crossing point of the two red dashed lines in Figure~\ref{fig:5}). (c) Same data, but scaled along vertical axis. The inset shows the scaling factor $a$ as a function of the background conductance at $B=0$. The red dashed line is a linear fit. The scaling factor $a$ is independent of $B$.}
    \label{fig:6}
\end{figure*}

The theoretical curves $G_\mathrm{vis}^\mathrm{th}$ obtained in this way do not match the experimental data $\Delta G$. At low temperatures the experimental and the theoretical values of conductance are close, but the experimental temperature dependence is weaker than the theoretically predicted one. Indeed, the hydrodynamic theory predicts $G_\mathrm{vis}^\mathrm{th} \sim 1/l_{ee} \sim T^2$ (neglecting the corrections $\sim~log(T)$), but the observed curves are much closer to linear dependence (see Appendix B for the details). 

At high temperatures the experimental curve $\Delta G$ decreases with increasing $T$ above a certain temperature. This decrease in conductance does not follow from the superballistic flow model, neither it can be explained by ohmic resistance of the 2DEG in series with the PC (ohmic resistance needed for that is more than order of magnitude above the total measured resistance of the system). Notably, for both electron densities at the maximum of $G_\mathrm{vis}$ the electron-electron mean-free path approaches $l_{ee}\approx\SI{1.0}{\micro m}$. If this experimental dependence was continued to higher temperatures, the peak would disappear at $T\approx\SI{20}{K}$ for high electron density, and at $T\approx\SI{12}{K}$ for low electron density. Both cases correspond to $l_{ee}\approx\SI{200}{n m} \approx d$ (see Appendix A).


This observation suggests the following explanation. At relatively low temperatures (i.e. $\SI{2}{K}<T<\SI{7}{K}$ for high electron density, Figure~\ref{fig:5}(b)) electron transport is mostly hydrodynamic and the electron-electron mean-free path is much longer than the width of the PC ($l_\tau \gg l_{ee} \gg d$). Consequently, the PC is the injection point for electrons, and all the interactions happen in the 2DEG outside of the PC. This corresponds to the superballistic flow model, with the predictions described in section IV.A. The conductance peak amplitude increases with $T$, but the functional dependence does not agree with the hydrodynamic prediction. Interestingly, a similar, approximately linear dependence was observed previously for the superballistic contribution of graphene PC conductance \cite{KrishnaKumar2017}. While the two results can not be compared directly due to the different methods used to extract the viscous contribution to the PC conductance, these observations hint that a more comprehensive model might be needed.

At high temperatures ($T>\SI{18}{K}$)  electron transport is still hydrodynamic, but the electron-electron mean free path becomes comparable to or smaller than the width of the PC ($l_\tau \gg d \gtrsim l_{ee}$). 
Electron-electron interactions now happen in the PC itself, and the electron transport tends toward Poiseuille flow through the channel rather than superballistic transport through the small PC. No conductance peak $G(B)$ is predicted for the hydrodynamic flow through the channel, and even the conductance peak from the classical size effect would be suppressed by hydrodynamic effects \cite{Scaffidi2017}. The PC in our experiment is not a long channel, but the top-gate defined potential provides a finite length of the PC which is more than its width $d$.

Consequently, even within the hydrodynamic model several regimes can be seen: superballistic flow is observed for lower temperatures, Poiseuille-like flow can be seen at higher temperatures. At intermediate temperatures a gradual transition between these two modes would be expected.

\subsection{Scaling}

The superballistic flow observed above (Section IV.A) can be additionally verified by examining $G(B)$ at different top-gate voltages $V_\mathrm{TG}$ at a fixed temperature and electron density. It can be shown that the curves can be scaled to collapse onto a single curve, and that the scaling parameter is close to the prediction of the hydrodynamic model. For these measurements we chose the temperature $T=\SI{4.2}{K}$ such that $l_\tau \gg l_{ee} \gg d$ and the model described in Section IV.A is applicable (Figure~\ref{fig:6}(a)).

Similar to the procedure used in the previous section, we separate the measured conductance into the V-shaped ballistic background and the hydrodynamic peak. Figure~\ref{fig:6}(b) shows the peak of conductance $G_\mathrm{vis}$ with the ballistic background subtracted for different values of $V_\mathrm{TG}$, while the inset depicts the ballistic background conductance $G_\mathrm{ball}$ at $B=0$ as a function of $V_\mathrm{TG}$. The peak amplitude changes with $V_\mathrm{TG}$, while its shape and width along the magnetic field axis are almost constant. Below we analyze this observation and compare it to the predictions of our model.

The amplitude of the peak explicitly depends on the effective width of the PC $d_\mathrm{eff}$ (Equation~\ref{eq:Gvis}), which explains the change of the amplitude with $V_\mathrm{TG}$. The constant shape of the observed peak can be demonstrated by scaling the curves at different $V_\mathrm{TG}$ according to 

\begin{equation}
    a(V_\mathrm{TG})G_\mathrm{vis}(B, V_\mathrm{TG}) = G_\mathrm{vis}^\mathrm{scaled}(B, V_\mathrm{TG}),
    \label{eq:scaling}
\end{equation}

where the scaling parameter $a$ depends only on $V_\mathrm{TG}$ and is chosen such that the mean square difference between the curves is minimal. Indeed, it can be seen that the scaled curves coincide (Figure~\ref{fig:6}(c)). The previously discussed model (Section IV.A) explains the constant width of the peak. The scaling behavior for the shape of the peak does not directly follow from the model; however the model can be used to study the behavior of the peak amplitude. 

Below we consider the simple approximation of the PC as a rectangular potential well, where $V_\mathrm{TG}$ affects only the effective width $d_\mathrm{eff}$ of the PC. In this case the ballistic conductance at zero magnetic field would be proportional to the PC width ($G_\mathrm{ball} \sim d_\mathrm{eff}$, follows from the Sharvin formula), while the viscous contribution $G_\mathrm{vis} \sim d_\mathrm{eff}^2$ (Equation~\ref{eq:Gvis}). Therefore, one can expect the relation $a \sim G_\mathrm{ball}^{-\gamma}$ between the scaling factor $a$ and the ballistic conductance $G_\mathrm{ball}$, where $\gamma=2$. The inset in Figure~\ref{fig:6}(c) shows the corresponding double logarithmic plot of the scaling parameter $a$ as a function of $G_\mathrm{ball}$. The curve is close to linear and the extracted $\gamma$ is between 1.5 and 1.8. This shows that the amplitude of the peak increases faster with $V_\mathrm{TG}$ than the ballistic background. In the argument above we ignored the dependence of the electron density in the PC on $V_\mathrm{TG}$ (which is considerable for the narrow PCs), so some discrepancy between the measured and predicted $\gamma$ is expected.

\section{Conclusions}

We performed measurements of conductance through point contacts in GaAs 2DEG and graphene at different temperatures, bulk carrier densities and magnetic fields. At elevated temperatures we observe a peak of conductance around zero magnetic field. The width of the peak along the magnetic field axis increases monotonically with temperature. The peak amplitude first increases and then (at least for GaAs devices) decreases with temperature. The shape of the peak also depends on temperature. For GaAs devices, the peak is particularly sharp (rounding on a scale of tens of $\SI{}{\mu T}$) at lower temperatures, $T\approx\SI{1.5}{K}$.

We propose a model based on a transition between different transport regimes and comparison of relevant length scales, which explains the observations qualitatively and, in many cases, quantitatively. For GaAs point contacts, the transition is from ballistic electron transport at very low temperatures (no magnetoconductance peak) to the few electron-electron interactions regime (small sharp peak) to hydrodynamic superballistic flow (amplitude of the peak increases) to hydrodynamic Poiseuille-like transport (the peak slowly disappears). For graphene point contacts, only the first and the third transport regimes are observed directly, although the last one should be possible at even higher temperatures than available in our experiment. 

\begin{acknowledgements}

The experimental data was discussed at an early stage with Leonid Levitov. His insights led to further experiments and to an intuitive understanding of the narrow magnetoconductance peak. We gratefully acknowledge these discussions.

We thank O. Zilberberg, A. Leuch, F.K. de Vries and H. Duprez for valuable discussions and comments and P. Märki, T. Bähler as well as the FIRST staff for technical support. We acknowledge financial support by the European Graphene Flagship Core3 Project, H2020 European Research Council (ERC) Synergy Grant under Grant Agreement 951541, European Union’s Horizon 2020 research and innovation programme under the Marie Sklodowska-Curie Grant Agreement No. 766025, Eidgenössische Technische Hochschule Zürich (ETH Zurich) and the Swiss National Science Foundation via National Center of Competence in Research Quantum Science and Technology (NCCR QSIT). K.W. and T.T. acknowledge support from the JSPS KAKENHI (Grant Numbers 19H05790 and 20H00354).

\end{acknowledgements}

\renewcommand{\theequation}{A.\arabic{equation}}

\renewcommand{\thefigure}{A.\arabic{figure}}

\revappendix
\setcounter{figure}{0}
\setcounter{equation}{0}

\section{Relevant length scales for GaAs}

\begin{figure*}
    \includegraphics[width=\linewidth]{{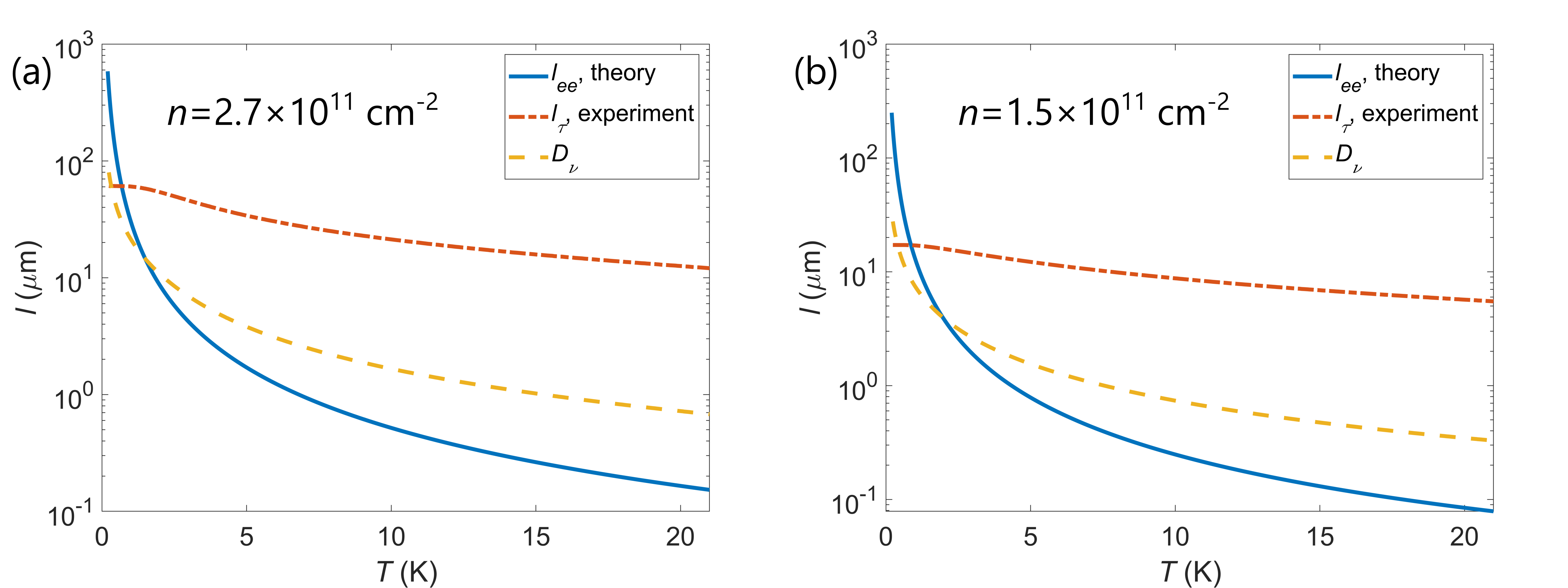}}
    \caption{Relevant length scales for GaAs as a function of temperature for electron densities $n=\SI{2.7e11}{c m^{-2}}$ (a) and $n=\SI{1.5e11}{c m^{-2}}$ (b). Solid blue line shows calculated electron-electron mean-free path $l_{ee}$, dotted-dashed red line shows measured transport (momentum-relaxing) mean-free path $l_\tau$, dashed yellow line shows viscous length $D_\nu$.}
    \label{fig:A1}
\end{figure*}

Transport mean-free path $l_\tau$, also known as momentum relaxing mean-free path, was extracted from standard mobility measurements in the wide Hall bar geometry. Viscous length $D_\nu$ was calculated according to \ref{eq:dnu}.

Electron-electron mean free path was calculated numerically according to \cite{Jungwirth1996}, assuming spin-independent scattering function in the random-phase approximation. The formulae for electron-electron scattering length $l_{ee}$ of a particle at energy $\tilde{\xi}_k = \frac{1}{k_BT}\left(\frac{\hbar^2k^2}{2m^*}-E_F\right)$ (dimensionless, relative to Fermi energy $E_F$) interacting with equilibrium Fermi sea at temperature $T$ were acquired by combining equations (7, 17-18, 32-33) of \cite{Jungwirth1996}:

\begin{widetext}
\begin{equation}
    l_{ee}(\tilde{\xi}_k, T) = \frac{\pi^2\hbar^4\sqrt{2\pi n^3}}{m^{*2}w^{fb}(k_BT)^2\left[1-n_F(\tilde{\xi}_k)\right]\left\{\frac{1}{2}\left(\pi^2+\tilde{\xi}_k^2\right)\left[\frac{\ln8}{2}+\ln\left(\frac{E_F}{k_BT}\right)\right]-F(\tilde{\xi}_k)\right\}\left[1+\exp(-\tilde{\xi}_k)\right]},
\end{equation}
where
\begin{equation}
    F(\tilde{\xi}_k) = \frac{1}{2}\left[1-n_F(\tilde{\xi}_k)\right]^{-1}\int_{-\infty}^{+\infty}d\tilde{\xi}_{k'}\int_{-\infty}^{+\infty}d\tilde{\xi}_p\ln\left|(\tilde{\xi}_{k'}-\tilde{\xi}_p)(\tilde{\xi}_k-\tilde{\xi}_{k'})\right|n_F(\tilde{\xi}_p)\left[1-n_F(\tilde{\xi}_{k'})\right]\left[1-n_F(\tilde{\xi}_k+\tilde{\xi}_p-\tilde{\xi}_{k'})\right],
\end{equation}
\begin{equation}
    w^{fb} = \frac{1+\left(1+\frac{1}{\sqrt{2}r_s}\right)^{-2}}{2}, r_s = \frac{m^*e^2}{8\pi\hbar^2\epsilon\epsilon_0\sqrt{\pi n}}
\end{equation}
$n_F(\tilde{\xi}) = \frac{1}{1+\exp(\tilde{\xi})}$ is a Fermi-Dirac distribution. The final result for electron-electron mean-free path follows from averaging of $l_{ee}(\tilde{\xi}_k, T)$ weighed by the derivative of the Fermi-Dirac distribution:
\begin{equation}
    l_{ee}(T) = \int_{-\infty}^{+\infty}l_{ee}(\tilde{\xi}_k, T)\left(-\frac{\partial n_F}{\partial\tilde{\xi}_k}\right)d\tilde{\xi}_k
\end{equation}
\end{widetext}

\section{Additional Measurements}

Additional measurements were performed to confirm the reproducibility of the results and provide further evidence for the suggested explanations.

\subsection{GaAs Point Contacts}

\begin{figure*}
    \centering
    \includegraphics[width=\linewidth]{{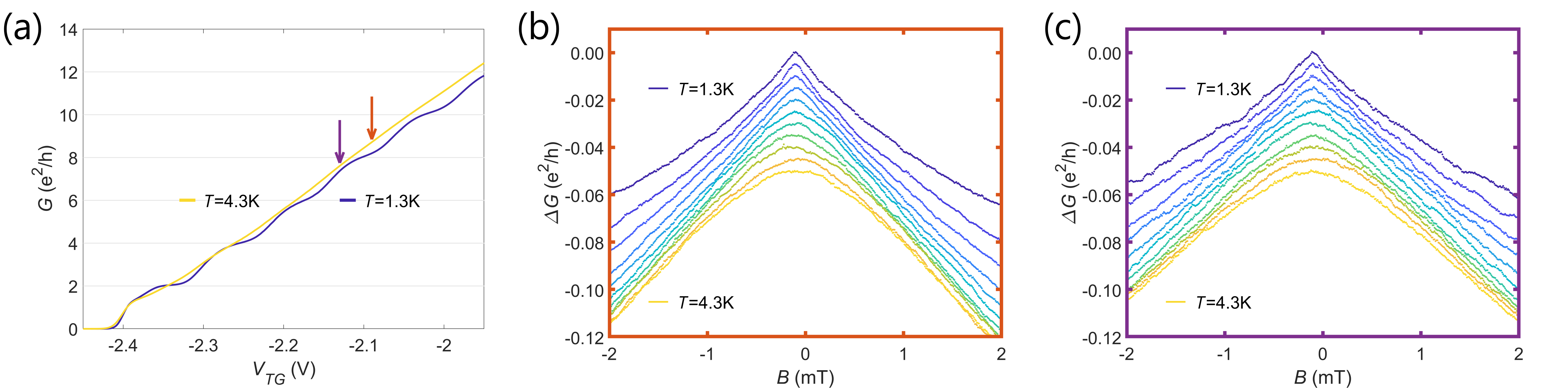}}
    \caption{(a) PC conductance $G(V_\mathrm{TG})$ for $n=\SI{2.7e11}{c m^{-2}}$ and two temperatures $T=\SI{1.3}{K}$ and $T=\SI{4.3}{K}$. Two arrows denote values of $V_\mathrm{TG}$ where the magnetoconductance measurements were performed, which are shown in (b) and (c) as $\Delta G(B,T) = G(B,T) - G(0,T)$. In the last two panels the curves are offset for clarity, the step in temperature between the curves is $\SI{0.3}{K}$.}
    \label{fig:E1}
\end{figure*}

\begin{figure*}
    \centering
    \includegraphics[width=\linewidth]{{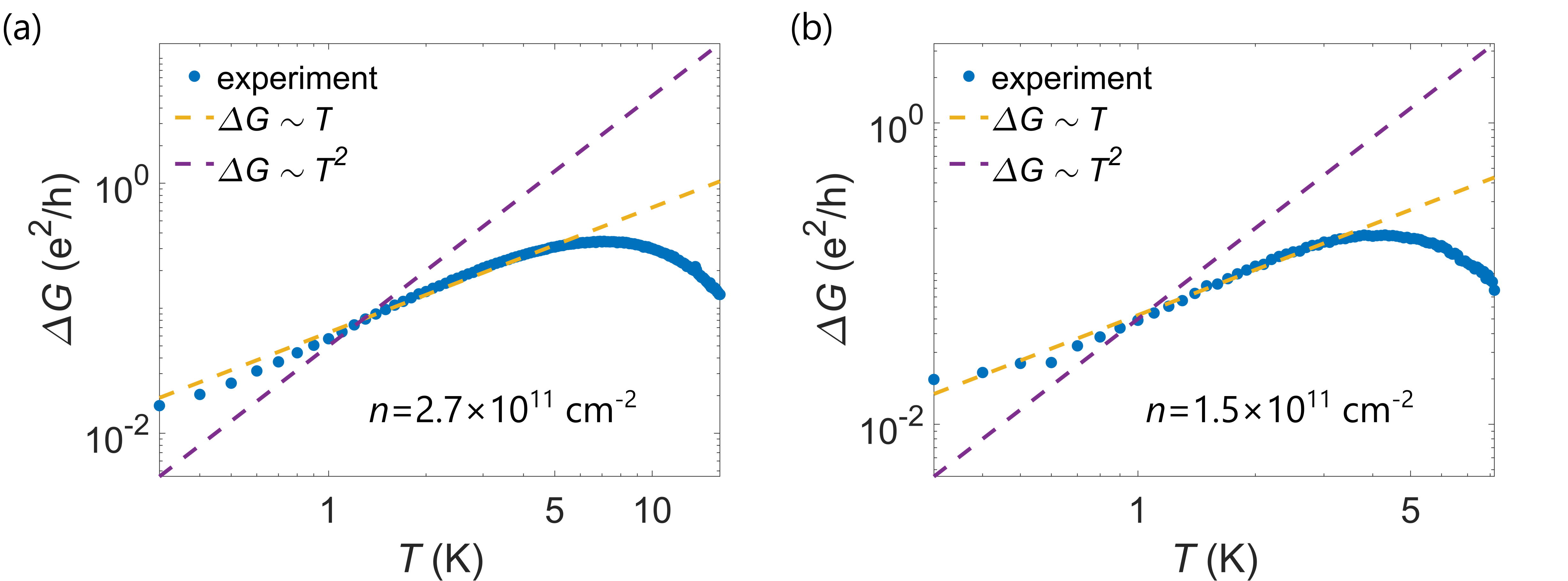}}
    \caption{Magnetoconductance peak amplitude $\Delta G$ (as defined in section IV.C) as a function of temperature shown in a double logarithmic scale for two electron densities $n=\SI{2.7e11}{c m^{-2}}$ and $n=\SI{1.5e11}{c m^{-2}}$. Dashed lines show linear and parabolic functions for comparison.}
    \label{fig:E2}
\end{figure*}

Figure~\ref{fig:E1}(a) shows conductance $G$ of the GaAs PC described in the sections II and III as a function of top gate voltage $V_\mathrm{TG}$ for $T=\SI{1.3}{K}$ and $T=\SI{4.3}{K}$. At low temperature, the standard conductance steps of $2e^2/h$ are observed. At high temperature, the conductance steps are much less pronounced and the overall conductance is increased more for higher values of $V_\mathrm{TG}$. Measurements are performed at $n=\SI{2.7e11}{c m^{-2}}$.

Examples of $\Delta G(B,T) = G(B,T) - G(0,T)$ in a narrow range of $B$ are shown in Figures~\ref{fig:E1}(b,c) for two different values of $V_\mathrm{TG}$ (one between the conductance plateaus, one almost on the plateau). The observed behavior is the same for these two cases.

In the section IV.C the dependence of the amplitude of the magnetoconductance peak on $T$ is discussed. Same data as in Figure~\ref{fig:5}(b,c) is shown in double logarithmic scale in Figure~\ref{fig:E2}(a,b). The observed curves at low temperatures are much better described by $\Delta G \sim T$ and not $\Delta G \sim T^2$ expected from hydrodynamic theory. $\Delta G$ is defined as in section IV.C.

\begin{figure*}[h]
    \includegraphics[width=\linewidth]{{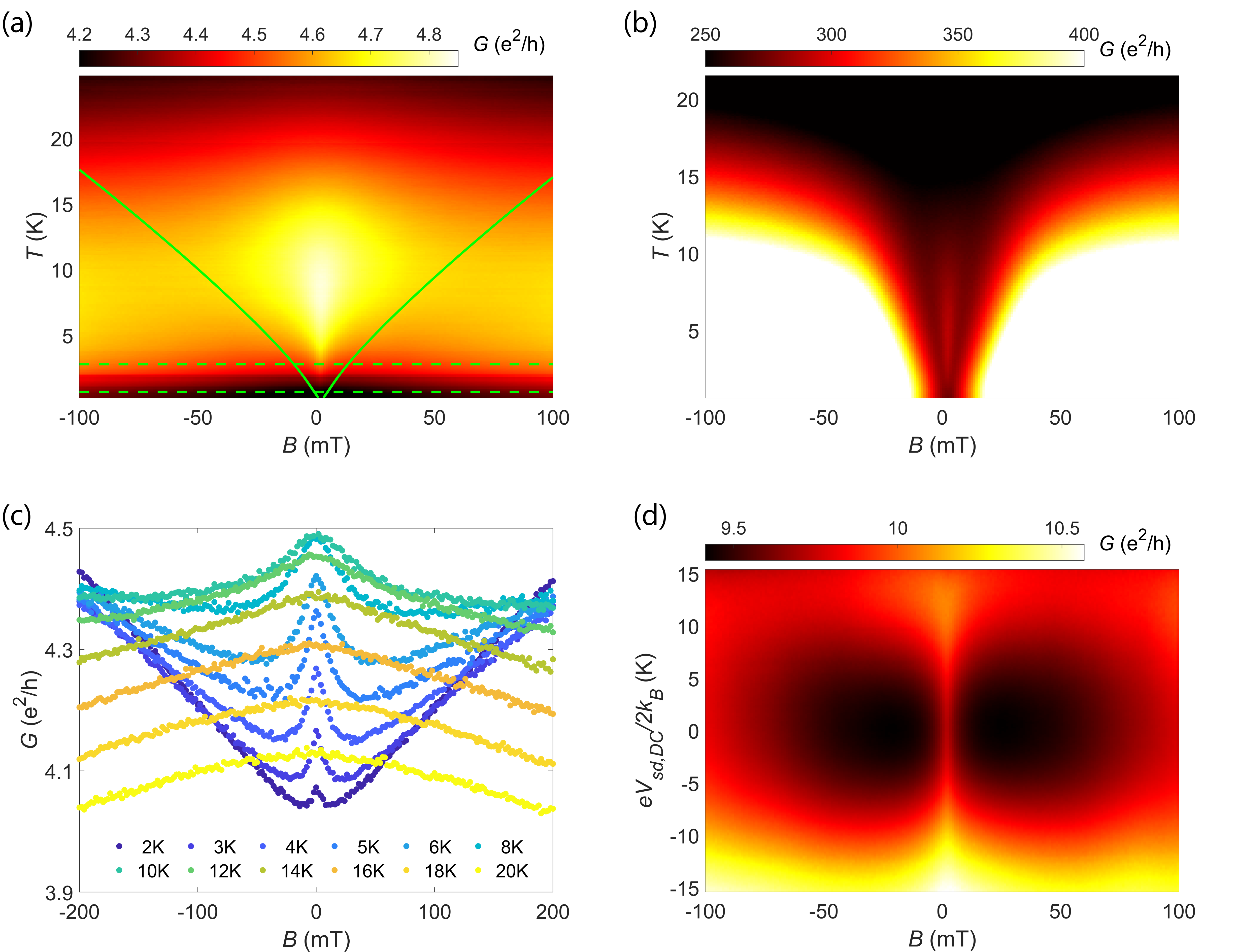}}
    \caption{(a) Conductance $G(B,T)$ of a second GaAs PC on the same wafer as described in Section II, electron density $n=\SI{2.7e11}{c m^{-2}}$, lithographic width $d=\SI{250}{n m}$. (b) Conductance $G(B,T)$ of a wide $d=\SI{4}{\mu m}$ GaAs PC on the same wafer. (c) Conductance $G(B,T)$ of a GaAs PC on a different wafer (electron density $\SI{1.8e11}{c m^{-2}}$, mobility up to $\SI{4e6}{cm^{2}/Vs}$, PC width $\SI{500}{n m}$). (d) Conductance $G$ as a function of magnetic field $B$ and DC bias voltage $V_{sd,DC}$ for the original PC (as described in Section II), measured at $T=\SI{4.2}{K}$.}
    \label{fig:A2}
\end{figure*}

Figure~\ref{fig:A2}(a) shows the conductance of a different PC of the similar dimensions to the one described in the sections II and III,  fabricated on the same GaAs wafer. Measurements were performed at electron density $n=\SI{2.7e11}{c m^{-2}}$. The observed $G(B,T)$ is qualitatively and quantitatively similar to the one presented in Figure~\ref{fig:2}(a).

Figure~\ref{fig:A2}(b) shows the conductance of a very wide constriction $d = \SI{4}{\mu m}$ (same GaAs wafer, density $n=\SI{2.7e11}{c m^{-2}}$). A weak magnetoconductance peak is present around $B=0$, although it is much less pronounced compared to the case of narrow PCs.

Similar conductance peak was observed in a different GaAs wafer (Figure~\ref{fig:A2}(c)). This wafer has no back-gate, electron density is $n=\SI{1.8e11}{c m^{-2}}$, low-temperature mobility is $\SI{4e6}{cm^{2}/Vs}$, PC width $d = \SI{500}{n m}$, 2DEG is $\SI{130}{n m}$ below the surface.

Instead of increasing temperature of the sample, higher bias voltage can be applied to achieve a qualitatively similar effect of broadening of the Fermi distribution. Figure~\ref{fig:A2} (d) shows conductance of the GaAs PC (wafer and geometry as described in Section II) as a function of $B$ and applied DC bias $eV_{sd,DC}/2k_B$. The measurements were performed at $T=\SI{4.2}{K}$, applied bias is a sum of a small AC signal for conductance measurement and large DC signal for increasing the effective temperature of the 2DEG. The magnetoconductance peak around $B=0$ is present for all applied values of the DC bias. The peak width is almost constant for $eV_{sd,DC}/2k_B < \SI{5}{K}$, where effective temperature of the 2DEG is mostly defined by the lattice temperature, and therefore almost constant. The peak width increases with the DC bias for $eV_{sd,DC}/2k_B > \SI{5}{K}$, where the effective 2DEG temperature is mostly proportional to the DC bias. This result is in agreement with the temperature dependencies $G(B,T)$ described above.

\subsection{Graphene Point Contacts}

\begin{figure}
    \includegraphics[width=.95\linewidth]{{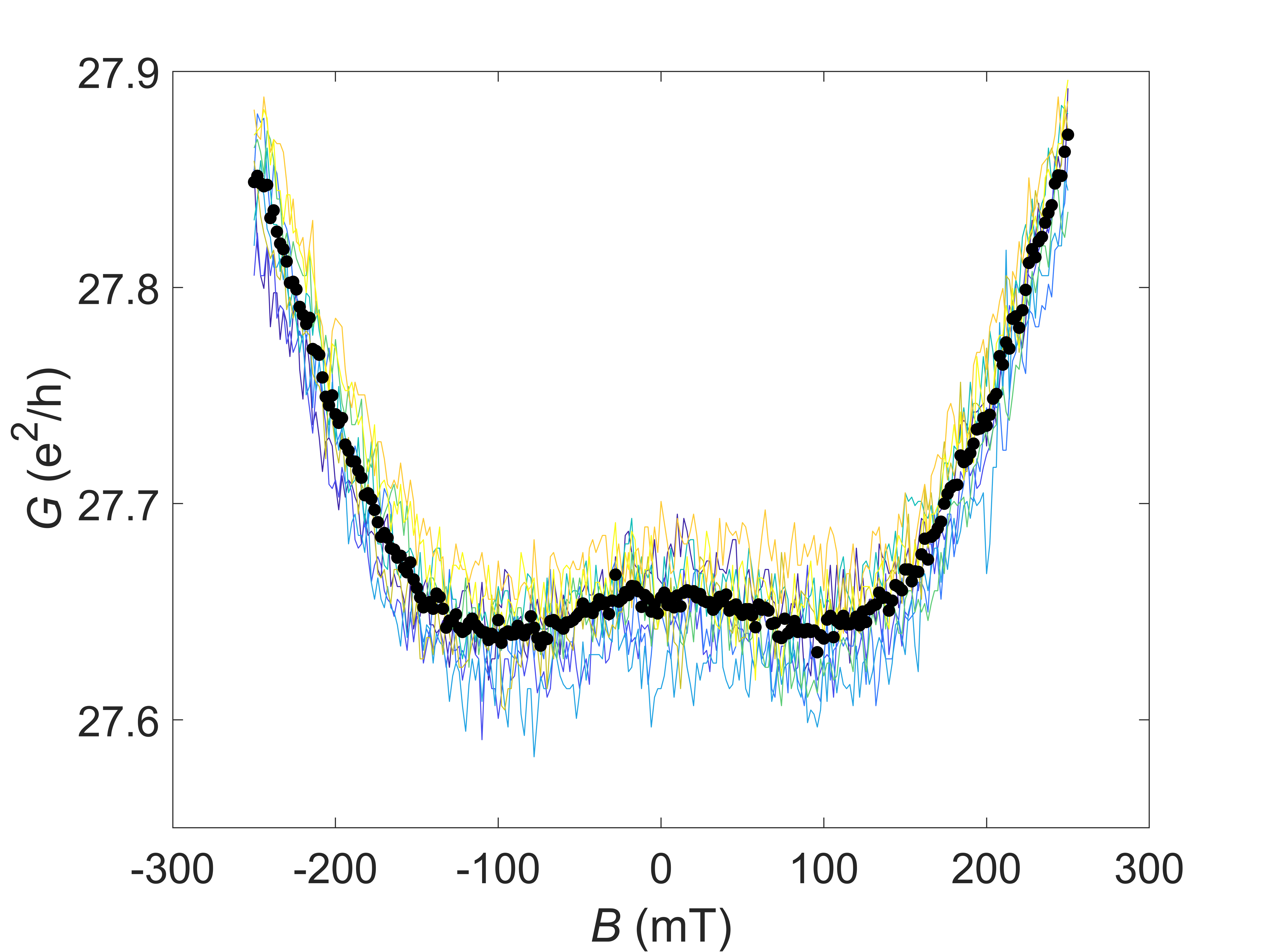}}
    \centering
    \caption{Conductance $G(B)$ of a PC in a bilayer graphene sample, $T=\SI{95}{K}$. The measurement was repeated nine times, each one shown in a narrow line of a different color. Black dots represent the average of the 9 measurements.}
    \label{fig:A3}
\end{figure}

Unlike monolayer graphene, bilayer graphene (BLG) devices allow gate defined structures and therefore better control of geometry. Bilayer graphene is also known to show hydrodynamic behavior \cite{Berdyugin2019}. Unfortunately, the gap in the bandstructure induced by vertical electrostatic displacement field in BLG is small compared to the bandgap of conventional semiconductors, such as GaAs. At temperatures high enough for hydrodynamic effects ($\approx\SI{100}{K}$) the current leak through the depleted region below the top-gates is significant and prevents proper confinement in the PCs \cite{HiskeThesis}. Regardless, we attempted to measure the magnetoconductance peak in the gate defined PCs in the BLG device. The data shown in Figure~\ref{fig:A3} is measured at $T=\SI{95}{K}$ and electron density $n\approx\SI{2e12}{c m^{-2}}$. The sample is sample I described in Appendix E of \cite{HiskeThesis}. A slight increase in conductance was observed around $B=0$. The effect is much weaker than in monolayer graphene devices and was only observed in a narrow window of parameters. In principle, higher displacement fields can produce higher bandgap and correspondingly smaller leakage currents at high temperatures, potentially improving the data quality. However, for our sample this was not possible due to a leak between gates and graphene at higher gate voltages. Further studies are needed in order to draw a clear conclusion regarding the magnetoconductance peak in a gate-defined BLG PCs.

\bibliography{main}

\end{document}